\documentclass[a4paper]{article}

\usepackage[english]{babel}
\usepackage[utf8x]{inputenc}
\usepackage[T1]{fontenc}
\usepackage{amssymb}

\usepackage[a4paper,top=3cm,bottom=2cm,left=3cm,right=3cm,marginparwidth=1.75cm]{geometry}

\usepackage{amsmath}
\usepackage{graphicx}
\usepackage[colorinlistoftodos]{todonotes}
\usepackage[colorlinks=true, allcolors=blue]{hyperref}

\title{Blending physical knowledge with mathematical form in physics problem solving}
\author{Mark Eichenlaub, Edward F. Redish}

\begin{document}
\maketitle
\begin{abstract}
For physicists, equations are about more than computing physical quantities or constructing formal models; they are also about understanding. The conceptual systems physicists use to think about nature are made from many different resources, formal and not, working together and inextricably linked. By blending mathematical forms and physical intuition, physicists breathe meaning into the equations they use, and this process is fundamental to what it means for an expert to understand something. In contrast, in physics class, novice students often treat mathematics as only a calculational tool, isolating it from their rich knowledge of the physical world. We are interested in cases where students break that pattern by reading, manipulating, and building equations meaningfully rather than purely formally. To find examples of this and explore the diversity of ways students combine formal and intuitive resources, we conducted problem-solving interviews with students in an introductory physics for life sciences class. During the interviews, we scaffolded student use of strategies which call for both formal and intuitive reasoning, such as ``examine the extreme cases'' and ``think about the dimensions''. We use the analytic framework of epistemic games to model how students used the strategies and how they accessed problem-solving resources, and we present evidence that novice students using these strategies accessed more expert-like conceptual systems than those typically described in problem-solving literature. They blended physical intuition with mathematical symbolic templates, reconceptualized the nature of variables and equations, and distinguished superficially-similar functional forms. Once introduced to a strategy, students sometimes applied it to new scenarios spontaneously or applied it in new ways to the present scenario, acknowledging it as a useful, general purpose problem-solving technique. Our data suggests that these strategies can potentially help novice students learn to develop and apply their physical intuition more effectively.

\end{abstract}

\section{Introduction}\label{intro}

Physicists and educators have long held problem-solving to be one of the key tools to help students understand physics\cite{meltzer2015brief}. If problem-solving is a bridge to expert-like understanding, we should find ways to let students experience expert-like thinking in as many dimensions as possible while working problems. This includes learning new physical concepts and mathematical techniques, because experts and novices differ greatly in the amount of physics and math they know. But experts also diverge from novices in their problem-solving strategies, their patterns of metacognition \cite{schoenfeld2016mathematical}, their epistemological stances towards their work (and abilities to negotiate between various stances), their conception of what mathematical entities are, and their expectations for how to derive meaning from their work. These differences between experts and novices are part of a ``hidden curriculum'' that students need to learn as they progress in physics, but which we rarely teach explicitly.\cite{redish2010introducing}

In particular, researchers have singled out math as a particular sticking point in problem solving in introductory physics. Much of the existing research seeks to document student understanding, or misunderstanding, of particular mathematical tools, such as differentiation or coordinate systems. Our teaching experience shows that even when students appear to have mastered the appropriate tools in previous classes, they may still struggle to use those tools effectively in physics problems. In previous work, one of us (Redish) \cite{redish2015language} laid out an argument that this is largely because the ways that physicists make meaning with mathematics are unfamiliar to students. Even if they are skilled with the manipulations of algebra and calculus, students' expectations about how to interpret variables may lead them astray. For example, many students, given a problem about test charges and electric fields, will say that changing the magnitude of a test charge changes the magnitude of the electric field it measures. They reason from the equation $E = F/q$ that if $q$ increases, $E$ decreases. The students understand the math involved well, but don't account for the way the force on a charge changes with the charge - there was a hidden functional dependence they did not see because physics culture assumes the reader will associate every symbol (in this case, $F$) to its physical meaning. That would make the functional dependence of $F$ on $q$ clear, but students don't yet expect to have to find this physical meaning when solving problems. The challenge for educators is to create problems and problem-solving environments that encourage students to search for physical meaning in mathematics.

In creating problems, educators often separate ``qualitative'' problems that test and build intuition from ``quantitative'' problems to develop mathematical skills \cite{hsu2004resource}, indicating an implicit assumption that these are separate faculties that are used and developed individually. We believe that for experts, intuition and mathematics are not insulated from each other or even cleanly separable. Instead, they reinforce each other; intuition is often connected to mathematics and mathematics is understood partially via intuition. While solving a problem, an expert will blend mathematical forms such as equations (or abstracted properties of equations), with intuitive conceptual schema to create richer mental spaces than those derived from formal mathematics alone.

For an example of what we mean, we look at Sherin's \cite{sherin2001students} description of ``symbolic forms'', a class of blended intuitive-formal conceptual structures that experts (and in Sherin's case, second-year physics students) use to understand equations. To introduce symbolic forms, we'll take an example from Sherin, who describes two students thinking about a ball falling through the atmosphere at terminal velocity. The students intuitively understand that air drag and gravity are both acting on the ball, but balance each other out, leaving no net acceleration. In Sherin's account, the students activate a conceptual schema for ``balancing'' of competing influences. This balancing schema could potentially match many different physical scenarios, or even everyday scenarios, such as expenses balancing out income when breaking even financially, but here is it called to understand air drag and gravity. The students then associate the balancing schema with the abstracted symbol template for equations, $\square = \square$, where each square represents one of the two balancing influences. The students know that they are looking for an equation with an expression related to gravity on one side and an expression related to air drag on the other. The students' work on a specific equation is then informed by this pairing of the intuition behind balancing with the symbolic template. The combined intuition and formal structure are collectively a symbolic form. Sherin identified 21 symbolic forms in his data corpus; our purpose here is to use them as one example of blended intuitive and formal thinking  that is found in experts and potentially in students as well.

Symbolic forms are not a complete account of how physicists make meaning with equations. The example of failed meaning-making in the equation $E = F/q$, cited earlier, involves the correct use of the symbolic form Sherin identified as ``prop-'', where a schema related to ``if one goes up, the other goes down'' is blended with the symbol template $\left[\frac{\ldots}{\ldots x \ldots}\right]$, but this symbolic form alone wasn't enough to lead students to the right answer.

We cannot give a full account of all the ways experts bring meaning into equations, but as a second example, we consider experts' ontology of equations, i.e., the types of objects equations are in experts' conceptual schemas. For example, here are a few examples of physicists writing about the relation between the Yukawa potential, $V(r) = \frac{qe^{-m r}}{r}$ and the Coulomb potential, $V(r) = \frac{q}{r}$.

\begin{quote}In the limit of $m\to 0$ the Yukawa potential becomes the Coulomb or gravitational potential\ldots\cite{Heile}\end{quote}

\begin{quote}...if we choose \ldots $m_0 = 0$, the potential reduces to the Coulomb potential energy\ldots\cite{townsend2000modern} [source uses $m_0$ in place of $m$] \end{quote}

\begin{quote}We can take the limit $\alpha \to 0$ and recover the Coulomb potential.\cite{hassani2013mathematical}[source uses $\alpha$ in place of $m$]\end{quote} 

\begin{quote} The Coulomb potential of electromagnetism is an example of a Yukawa potential \ldots \cite{wiki:xxx} \end{quote}

\begin{quote} We see \ldots that if the mass $m$ of the mediating particle vanishes, the force produced will obey the $1/r^2$ law. If you trace back over our derivation, you will see that this comes from the fact that the Lagrangian density for the simplest field theory involves two powers of the spacetime derivative \ldots \cite{zee2010quantum} \end{quote}
%
%
In some cases, physicists see themselves as enacting a change in the Yukawa potential. They or their reader actively ``take the limit'' or ``choose $m=0$''. Other times, the Yukawa potential changes, but there's no clear agent involved. It may ``become'' or ``reduce to'' the Coulomb potential and the mass may ``vanish'', but no entity is identified as enacting the change. In contrast to these dynamic descriptions, the relationship can also be described statically. Nothing in particular is happening when the Coulomb potential "is an example of" the Yukawa potential. 

This is just a sampling of physicists' language on the topic. The details of how they describe the Yukawa potential-Coulomb potential relationship may depend on both the physicist and the context of what they're communicating in complicated ways. Our goal here is simply to illustrate that there is a significant diversity of ways to conceptualize of an equation.

These examples come from professional, graduate, and upper-division undergraduate material, where such a diversity of conceptualizations of equations is commonplace. By contrast, in introductory physics textbooks, equations are usually  treated as static entities to be scrutinized.

\begin{quote}Outside the nucleus the nuclear force is negligible, and the potential is given by Coulomb's law, U(r) = +k(2e)(Ze)/r,\ldots \cite{tipler2007physics} \end{quote}

\begin{quote}Coulomb's law can be written in vector form (as we did for newton's law of universal gravitation in Chapter 6, Section 6-2), as $\mathbf{\vec{F}}_{12} = k \frac{Q_1Q_2}{r^2_{21}}\hat{\mathbf{r}}_{21}$ \ldots \cite{giancoli2000physics}\end{quote}

\begin{quote}The electric force acting on a point charge $q_1$ as a result of the presence of a second point charge $q_2$ is given by Coulomb's Law: $F = \frac{kq_1q_2}{r^2} = \frac{q_1}{q_2}{4\pi\epsilon_0 r^2} \cite{hyperCoulomb}$\end{quote}

The main exception we have observed is for descriptions of formal operations on them that come up during derivations (e.g. ``differentiate with respect to $t$'', ``set them equal to each other'', etc.), although the equations are also sometimes described as active entities, for example "Coulomb's law describes a force of infinite range which obeys the inverse square law \cite{hyperCoulomb}" in that they "describe" things, but this does not represent the same diversity of conceptions we saw with regard to the Coulomb and Yukawa potentials.

This mostly-static view of equations stands in contrast to introductory physics sources' descriptions of the physical quantities the equations represent

\begin{quote}
We can divide up a charge distribution into infinitesimal charges \ldots \cite{giancoli2000physics}
\end{quote}

\begin{quote}
The force exerted by one point charge on another acts along the line joining the charges. It varies inversely as the square of the distance separating the charges and is proportional to the product of the charges. \cite{tipler2007physics}
\end{quote}

In describing the force, field, or charges associated with Coulomb's law, introductory use both agentive language ("We can divide") and non-agentive ("The force \ldots acts\ldots"). The second quotation here also mixes dynamic ("varies inversely\ldots") with static ("is proportional to\ldots") language in the same sentence. So while a diversity of ontological viewpoints are generally considered acceptable for thinking about physics in introductory settings, this seems to apply much more to physical quantities than to equations, but as we move to more expert settings, the equations themselves take on the same diversity of ontologies.

Sfard's \cite{sfard1991dual} notion of conceiving of functions as either objects or processes is similar to ours, but here we consider ``process'' views where the equation itself is changing, as opposed to Sfard's notion of a static function which describes change when inputs transform into outputs. Our point here is simply to illustrate one more small piece of the diversity in expert conceptual systems used to make mathematics physically meaningful. This piece, like symbolic forms, is never explicitly taught. It is a part of the hidden curriculum, and something we can try to find evolving in students as they progress towards expertise.

Based on an exploratory analysis of problem-solving interviews, we suggest that students, in the right circumstances, use a large and diverse arsenal of productive, sophisticated, and creative ways to conceptualize physics problem-solving. They do not always access these resources when they would be productive, and many of the difficulties students experience with using math in physics are not so much difficulties of having the appropriate tools, but of applying them appropriately. While much of the hidden curriculum will need to be learned via years of enculturation in the physics community, there are entire swaths of it that don't need to be explicitly taught so much as activated. Small interventions that encourage students to use specific problem-solving strategies, can, in some cases, greatly enhance students' access to productive ways of thinking about mathematical tools that are rarely explicitly taught. 

The strategies we're investigating are commonplace, well-known to physicists, and generally well-regarded components of effective problem-solving. They include examining special and extreme cases, dimensional analysis, and estimation. Our contribution to understanding these strategies is to suggest that their scope can be very broad; they can be used at different stages of problem-solving and in different ways - and to give examples of how students using these strategies construct meaning from mathematical expressions in ways similar to how experts do it.

\section[Theoretical Framework: Resources, Framing, and Epistemic Games]{Theoretical Framework: \\ Resources, Framing, and Epistemic Games}

Our analysis is situated in the resource model \cite{Hammer2000,redish2004theoretical}. In this framework, students don't have monolithic conceptual understandings; they have many small pieces of knowledge, or resources that they can call on while solving a problem. When solving a problem, students will activate various resources and construct a solution based on them. If students don't solve a problem correctly, it may be that they don't have the appropriate resources, or that they do, but aren't activating them in that context. In the previous example of a test charge and the measured electric field, students did activate resources relating to understanding inverse mathematical relationships (including the prop- symbolic form), but did not activate resources related to the functional dependence of force. Whether or not students activate a resource can depend on how they associate it with other resources they are using, so in a future problem, students might improve their performance if they've learned to activate resources related to functional dependence when they see questions about forces in electromagnetism.

The issue is not so simple, though. The students in question were all able to recite the mantra ``the electric field is independent of the test charge''. In this sense, they knew the answer to the problem, but they didn't call on this knowledge, or if they did, didn't apply it. In addition to resources related to manipulating mathematical equations and resources related to intuitive understanding of physics, students also have ``epistemological resources'', resources related to how they seek to obtain and justify knowledge \cite{hammer2003tapping}. 

A student who uses an equation because it makes intuitive sense may come to the same answer as a student who uses an equation they found in a textbook they consider authoritative, but the way they are thinking about knowledge is very different; they are using different epistemological resources. The students who answer the test charge problem incorrectly are probably not activating epistemological resources related to interpreting each variable physically, or resources related to finding concordance between memorized facts (such as the electric field being independent of the charge) and the results of reasoning based on equations.

To understand why students sometimes use one set of epistemological resources and sometimes another, we use the lens of epistemological framing \cite{bing2009analyzing}. Because we could potentially use any resource at our disposal (i.e. every fact, technique, or type of reasoning we can conceive of) on a given problem, the space of problem-solving strategies we have to search through to find one effective approach is extremely large. We begin by narrowing the problem down to a certain type of problem, and then search through the resources we associate with that type. Calling on a physical principle to solve a problem requires activating different epistemological resources than using an equation does, and those resources often are associated with different epistemological framing. \cite{gupta2011beyond,kuo2013students} Students who answered that changing the magnitude of a test charge changes the magnitude of the measured electric field may have entered a ``calculation'' frame, and didn't remember or pay attention to their knowledge that the electric field is independent of the test charge because they didn't frame the task as one in which physical principles are relevant.

Moving towards expertise in problem solving is as much about using what resources you have effectively as it is about picking up new resources. As students work physics problems, they need to learn not only new content, but new ways of relating to the content. They need to be able to effectively frame epistemologically and activate appropriate resources. All of these are difficult tasks that live mostly in the hidden curriculum.

Analyses of problem solving often break the task down into a series of steps. Sometimes this is prescriptive, as when textbooks list a series of steps to make in solving a problem. For example, Redish \cite{redish2010introducing} describes a textbook with the following scaffold for problem solving
\begin{quote}
Model! - Make simplifying assumptions.

Visualize! - Draw a pictorial representation.

Solve! - Do the math.

Assess! - Check your result has the correct units, is reasonable, and answers the question
\end{quote}
and gives an example where the method failed. The textbook posed a question asking us to find the volume occupied by the water evaporated after sweating during exercise. The solution manual followed each step, finding that the volume was simply the volume of an ideal gas with the appropriate number of molecules, ignoring that the evaporated water will, by convection and diffusion, spread out over a very large volume. The textbook's solution manual follows each individual step, but nonetheless comes to a nonsensical answer to a problem by failing to ``tell the story of the problem''. From this example, Redish finds 
\begin{quote}Tying the analysis to a rubric – a formal set of mapped rules \ldots does not help if it does not also activate an intuitive sense of meaning by tying the problem to all we know and recognize about a system 
\end{quote}

We also view problem-solving as a series of steps, but not as steps for students to follow, but as a framework for researchers to understand how students solve problems. This approach is common in physics education research. For example, in analyzing student difficulties using math in physics, Wilcox et. al \cite{Wilcox2013} proposed the ACER framework, which consists of Activation of the tool, Construction of the model, Execution of the mathematics, and Reflection on the result.

Whereas a prescriptive problem-solving script tells students to follow precise steps in a given order, Wilcox et. al. write, ``\ldots we are not suggesting that all physics problems are solved in some clearly organized fashion, but a well articulated, complete solution involves all components of the ACER framework.'' That is, having the framework allows the researchers to narrow their focus and identify specific tasks students are struggling with, rather than simply bemoaning that they can't apply math appropriately. In that paper, Wilcox et. al. found that students' resources for the technique of taking a Taylor expansion weren't activated by the appropriate signal, which was one variable of interest being very much smaller than another, and suggested that problems be written to focus on building this particular association for students between signal and mathematical technique.

Frameworks like ACER are effective at picking out specific technical steps that students don't take in problem-solving. Our interest here is broader, including student epistemologies, attitudes towards mathematics, conceptualization of the entities involved, and other aspects of the hidden curriculum. The framework of epistemic games is a flexible one that allows analysis of both problem-solving moves and the motivations behind them.

We have previously discussed epistemological frames in problem-solving. Framing is a general feature in psychology, and when we work in a particular frame it often cues a script for how that type of activity typically goes, which sets expectations for what will happen next and what sorts of actions are appropriate \cite{goffman1974frame}. 

An epistemic game is a script that allows us to understand the moves students make in problem solving \cite{tuminaro2007elements}. As we watch students solving problems, we assign their problem-solving to some particular epistemic game, which we take to structure the types of resources they call on and the order in which they use them. An epistemic game will generally have a particular epistemological frame associated with it, but adds additional structure. The viability of epistemic games as an analysis framework stems from its psychological plausibility via the connection to psychological scripts and that, when Tuminaro and Redish \cite{tuminaro2007elements} analyzed student problem solving, they found that certain epistemic games were repeated many times on different problems and in different circumstances. The term ``epistemic game'' comes from Collins and Ferguson \cite{collins1993epistemic}, although the version we use here is that of Tuminaro and Redish \cite{tuminaro2007elements}.

In an epistemic game, as in games like solitaire or chess, one or several players make moves. These moves might be mathematical moves, such as \emph{add the same quantity to both sides of the equation}, conversational moves, such as \emph{offer a reason supporting your position}, or physical moves, such as \emph{draw a picture of the situation}. Because players can make various types of moves, analyzing the moves lets us focus on different aspects of the hidden curriculum in problem-solving. 

As the players of an epistemic game make moves, they gradually fill out an epistemological form, a template for what the solution to the problem should look like, which may be physical or verbal. Finally, players either reach the e-game's stopping condition and decide they are done, or else switch to a different game or give up on the solution attempt.

Tuminaro and Redish identified six common games that students play during problem solving, such as \emph{recursive plug-and-chug}, in which students identify a formula and put values into it without interpreting the results, and \emph{mapping meaning to mathematics}, which describes the  problem solving process in which students analyze the physics of a situation, turn their analysis into equations, manipulate the equations, and then turn the result into a new physical understanding.

Students use e-games to guide their inquiry, and their (generally unconscious) choices for what e-game to play have large effects on their problem-solving process. Different games have different rules about what sort of evidence is salient, what sort of moves are allowed, what type of arguments to give, and what it means to be done with a problem. When students get stuck on a problem or come to answers that don't make sense from the viewpoint of experts, they often have resources that would allow them to solve the problem, but never access them because they are not included in the current frame \cite{tuminaro2007elements,bing2012epistemic}. 

We do not consider playing an epistemic game favorable or unfavorable; that depends on which epistemic game and how appropriate it is to the situation. Epistemic games also aren't confined to students; experts play them as well, and do it very effectively. For example, in his short paper ``A Model of Leptons'' \cite{weinberg1967model}, Weinberg searches for an equation to describe leptons and their interactions. The method is to list various properties the equation should have---what symmetries it has, what types of solutions to avoid, etc. Each such consideration can be translated into a particular feature that the final equation should have, and by combining a sufficient number of features, only one equation is left that satisfies them all---the final equation derived for leptons and their interactions. Weinberg is playing an epistemic game we call ``significant features''. This is a game used to generate solutions to a given problem (as opposed to evaluating a proposed solution). To play, one lists relevant significant features a solution ought to have, such as a maximum at a certain place, or matching a certain symbolic form. Each feature is translated into a formal constraint or piece of the sought solution, such as the derivative being zero at the maximum or a symbolic template which matches the symbolic form appearing in the equation. As the player discovers more features and their associated forms, they gradually fill out the equation (or plot or other form) they are seeking. The game ends when they either decide they have completely specified the answer to the problem or decide that they don't know enough features to do so.

In Sherin's work \cite{sherin2001students}, two students decide that under constant acceleration the equation for velocity as a function of time is either $v(t) = v_0 + at$ or $v(t) = v_0 + \frac12at^2$, but cannot decide between the two. Sherin analyzes this as using the ``base plus change'' symbolic form. Students conceptualize the situation as velocity starting at some given value, then changing to a new value, and realize that this maps onto the symbolic template $\square + \Delta$. The symbolic form doesn't distinguish between the terms $at$ and $\frac12 at^2$ as ``changes'' to map onto $\Delta$ in the symbolic template. Both are positive (for positive acceleration and time) and indicate an object speeding up. Sherin's analysis is that using only a symbolic form isn't enough for students to determine the correct equation. We agree, and add that the students are playing the same ``significant features'' game that Weinberg did in building a model of leptons. They begin with a feature they want to the solution to have - matching the conceptual schema of base + change, and translate that into a mathematical form - the $\square + \Delta$ symbol template. Although they ran out of features to finish constraining their answer to the one correct answer, they were nonetheless playing the same epistemic game, just with very different material and at different levels of expertise.

\section{Data and Analysis}

The students we interviewed were enrolled in an introductory physics for life science course at the University of Maryland. Most are juniors, with some sophomores and seniors. The course prerequisites include one semester each of calculus, probability, chemistry, and two semesters of biology. Students are mixed between having taken physics in high school and not.

This is a population of relative novices in physics, but who have taken from 5 - 12 college science courses before taking this one; they generally have strong expectations about how science courses and problem-solving in them work, which the instructor (Redish) routinely challenges.(See \cite{redish2014nexus} for more details on the creation and principles behind the course.) We conducted hour-long interviews with students enrolled in the first of two semesters of this course. All interviews used a think-aloud protocol, encouraging students to write and articulate their thoughts at all times as they solved problems. Some interviews were one-on-one with the interviewer (Eichenlaub) and other were group interviews in which the interviewer was present but participated minimally, with occasional small interventions designed to prompt use of specific problem-solving strategies. 

With these interviews, we were interested in the breadth of approaches and conceptualizations students take in problem solving, including whether and how they blend physical intuition with mathematical formalism and how they conceive of variables, parameters, and entire equations. We chose problems and problem-solving strategies that we hoped would elicit epistemic games with a strong interplay of intuition and formalism in hopes of bringing out a diversity of interesting conceptual systems in students' solution attempts. The strategies we investigated were \textit{examine extreme or special cases}, \textit{dimensional analysis}, and \textit{estimation}, chosen especially because they are all familiar parts of an expert physicists' toolkit, but are not always taught explicitly at the introductory level.

We wanted to make fine-grained analysis of small, interesting incidents in our interviews, so we took video of the interviews ensuring that the field of view captured all students (for group interview) or student and interviewer (for one-on-one interviews) so that we could reference speech, gesture, and other expressions. Students wrote on a whiteboard, which we photographed at the interview's conclusion.

Our goal in analyzing these interviews was to generate hypotheses about cognitively-rich ways that students can interact with math and physics. This was exploratory analysis, not confirmatory, so the results we present here are case studies to be examined in more detail in the future. Our focus was on finding particularly interesting moments throughout the problem-solving sessions, including moments of blended mathematical/intuitive sensemaking and moments that show how students conceive of the mathematical entities they're working with. To that end, we reviewed the videos highlighting incidents that stood out to us, then discussed them together to generate hypotheses regarding student conceptualizations that interested us. Here we present those hypotheses along with descriptions of the incidents that we watched while generating them.

Below, we describe each strategy and report briefly on how students in our interviews took up the strategy before discussing, through the lens of epistemic games, specific cognitive aspects of problem-solving that these strategies elicited.

\subsection{Extreme and Special Cases}

Most physical systems we examine in problem solving have one or more free parameters that enters the problem. For example, in trying find the effective spring constant of two springs connected in series to form a single combined spring, the individual spring constants are such parameters. If we set one of these parameters to its largest or smallest possible value, we're looking at an extreme case. So for springs in series, we could set the second spring constant to be infinite, in which case it is completely rigid, does not contribute at all to the stretching of the combined spring, and the effective spring constant would simply be that of the other spring. Using this fact to try to understand something about the general situation is a strategy we call ``extreme case'' reasoning. We might also consider the case where the two spring constants are equal. Then each spring stretches the same amount, the total stretch is twice as much as the stretch of an individual spring, and the effective spring constant is half that of an individual spring. We call this ``special case'' reasoning. The two are almost the same, but extreme cases have been discussed independently in the literature, so we identify them as separate but closely-related reasoning strategies.

Clement \cite{clement2009extreme} studied extreme cases in a grade school setting, finding that looking at the extreme case helps students build vivid, dynamic mental imagery, consistently leading to better intuitive understanding of physics scenarios. Used in quantitative problem solving, extreme cases not only boost our intuition, but also allow us to connect that intuition to equations we've generated or are considering. Our accuracy and intuition for thinking about extreme cases has led physicists to make their study a standard problem-solving tool \cite{morin2008introduction}. Nearing \cite{nearing2003mathematical} elaborated on why extreme cases lead to better intuition in his undergraduate textbook on mathematical physics
\begin{quote}
How do you learn intuition?

When you've finished a problem and your answer agrees with the back of the book or with your friends or even a teacher, you're not done. The way to get an intuitive understanding of the mathematics and of the physics is to analyze your solution thoroughly. Does it make sense? There are almost always several parameters that enter the problem, so what happens to your solution when you push these parameters to their limits? In a mechanics problem, what if one mass is much larger than another? Does your solution do the right thing? In electromagnetism, if you make a couple of parameters equal to each other does it reduce everything to a simple, special case? When you're doing a surface integral should the answer be positive or negative and does your answer agree?

When you address these questions to every problem you ever solve, you do several things. First, you'll find your own mistakes before someone else does. Second, you acquire an intuition about how the equations ought to behave and how the world that they describe ought to behave. Third, It makes all your later efforts easier because you will then have some clue about why the equations work the way they do. It reifies the algebra. 
\end{quote}

Extreme cases, to Nearing, are not about the physics situation alone or the mathematical expression alone, but a way of bridging the two into a unified qualitative and quantitative understanding of physics.

In a prototypical use of the extreme or special case reasoning, students first derive an expression, in terms of parameters of the problem, that is a potential solution to the problem, for example finding the acceleration of a block in terms of various masses, angles, and coefficients of friction involved. They then use their physical intuition for extreme cases to evaluate this potential solution. 

This evaluative use can be analyzed as a ``sanity check'' epistemic game. This game begins after students generate a candidate solution to a problem, and is used to test whether the solution makes sense. The prototypical moves of the game are

\begin{enumerate}
\item{Identify a feature which the candidate solution intuitively ought to have.}
\item{Check whether the candidate solution has this feature.}
\item{If it does, identify a new feature the solution ought to have. If it does not, either reject the solution and start over, or enter a new epistemic game to determine why the solution and feature do not match.}
\item{Continue playing the game until you can't think of any more features or are satisfied with your confidence in the candidate solution.}
\end{enumerate}

When playing the sanity check game with the extreme case strategy, these moves could look like this:

\begin{enumerate}
\item{Identify a physical variable in the problem.}
\item{Imagine it becoming extremely large, extremely small, or some special value that stands out.}
\item{Intuitively identify the behavior of the system in this case.}
\item{Analyze the same limit of expression in the potential solution.}
\item{Compare the results of (3) and (4) for consistency. If they are consistent, confidence in the solution increases. If they are inconsistent, choose a new e-game to figure out whether it is your intuition or the mathematical expression that is incorrect.}
\item{Repeat for other variables in the problem.}
\end{enumerate}

This game encourages students to repeatedly compare a mathematical expression with a physical intuition, and so promises to be a good place to learn about how students use math to inform physical understanding and vice versa.

Although we've outlined a canonical version of the game above, physicists use extreme cases in many other ways. The snippets from physicists discussing the relation between the Yukawa and Coulomb potentials in section \ref{intro} discuss sending a parameter ($\alpha$) to an extreme (zero), but instead of examining the physical behavior of a system in this limit, they discuss an equation itself simplifying to a different equation. 

Further, in many cases beyond the introductory classroom, we can only find analytic solutions for the limiting cases of an equation, so studying the asymptotic behavior of otherwise intractable physical systems has become the most common analytical approach in modern mathematical physics \cite{bender1999advanced}. As a result, extreme cases and special cases lead to a host of useful tools, resources, and intuitions for physicists, including for example perturbation theory and the WKB method. The power of this game is one of the reasons that the predilection of introductory students to ``put numbers in right away'' (thereby reducing the problem to one that looks more like ``just math'') is so counter-productive.

\begin{figure}
\includegraphics[width=400px]{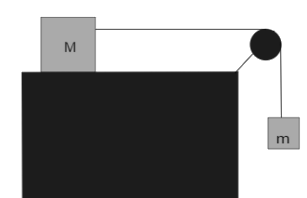}
\caption{The half-Atwood problem: A block of mass $M$ is attached to a block of mass $m$ via a massless string strung over a pulley as shown. The setup is frictionless. What is the acceleration of the block $m$?}
\label{fig:atwood}
\end{figure}

In interviews, we gave students several problems where we expected the extreme cases game to be useful: the half-Atwood machine (Figure \ref{fig:atwood}), the electric field on the axis of a ring of charge, springs in series and parallel, and the area of an ellipse.

\begin{figure}
\includegraphics[width=400px]{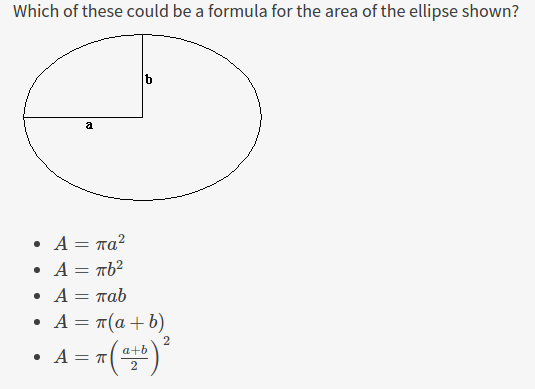}
\caption{The ellipse problem}
\label{fig:ellipse}
\end{figure}

In every case, we found that students have strong and accurate physical intuitions for the extreme or special cases. In some circumstances, students consistently spontaneously play the sanity check game using special case reasoning. For example, every student interviewed on the ellipse problem (Figure \ref{fig:ellipse}) considered the special case $a=b$, a circle, and used it to evaluate the given answers. No students, on the other hand, spontaneously checked the extreme case $a \to 0$, however, when prompted by the interviewer to consider ``a long, skinny ellipse'', most did use this extreme case to answer the question correctly.

Extreme/special case reasoning also proved consistently valuable to students answering the half-Atwood problem (Figure \ref{fig:atwood}) and to students finding the electric field on the axis of a ring of charge. 

The students in our interviews found this strategy less effective when asked to determine the effective spring constant of two springs connected in series. Asked to consider this problem without being prompted to think of extreme cases, Lizzie, Myra, and Lelia (pseudonyms) had the following discussion:

\begin{quote}
\begin{enumerate}
\item{Lelia: What's Hooke's law again? Oh yeah, T is this. [writes an equation for Hooke's law] So in this. The length would technically be twice as long.}

\item{ Lizzie: oh for the two}

\item{ Lelia: technically this k coefficient would be twice as long as one of them.}

\item{ Lizzie: yeah [erases board and writes $T = k \Delta L$]}

\item{ Lelia: so I think k-series would be them added together. Cause I remember I remember from}

\item{ Lizzie: the homework}

\item{ Lelia: yeah there's two connected the new k coefficient is twice as much, I think.}

\item{ Lizzie: we have two k's. [all writing equations involving $k$, $T$, and $\Delta L$]}

\item{ Lizzie: k-series would be k-one plus k-two}

\item{ Lelia: yeah, that's what I'm thinking}
\end{enumerate}
\end{quote}

Lizzie, Lelia, and Myra (did not speak above) associate higher spring constants with more length of the spring, leading them to conclude that springs in series have an a spring constant that adds. After working on other problems for twenty minutes, they returned to the springs, and the interviewer asked what would happen if one spring were much stiffer than the other

\begin{quote}
\begin{enumerate}
\item Lizzie: the stretch, the easy one would stretch a lot

\item Lelia: and the hard one would stretch a little bit, so the total stretch would be mostly due to the softer spring. so i mean again I guess k-constant would be the softer one.

\item Lizzie: but the hard one would still contribute a little bit

\item Lelia: yeah, but we don't know. I don't know how much, you know what percentage

\item Myra: can we like divide it by the number of springs?

\item Lelia: like k-one plus k-two divided by two or something?

\item Lizzie: or n?

\item Myra: cause I'm thinking because if one is way easier to stretch and the other one is not stretching at all, but each spring is still contributing some stretching, so then you divide it by the number of springs.
\end{enumerate}
\end{quote}

Their physical intuition is correct, but in the remaining time, they are unable to match their intuition to an equation, and ultimately revert to their original answer of $k_{eff} = k_1 + k_2$. Although their effort to play extreme cases didn't result in a correct equation, they did make correct conclusions about the mathematical form of the answer, specifically that the effective constant should be (very nearly) the same as that of the softer spring, and they consistently attempted to match physical intuition to equations. However, without a clear mapping from spring constants onto physical stiffness, it was difficult for them to find a correct equation.

\subsection{The Dimensional Analysis Game}

There are several strategies based on the idea that if two physical quantities are equal, they must have the same dimensions. We refer to these strategies collectively as ``dimensional analysis'', and they are taught extensively at the introductory level \cite{Robinett2015}, while also remaining of professional interest to physicists for more than a century \cite{bridgman1922dimensional}. A prototypical example of playing the sanity check epistemic game for evaluating a formula using dimensional analysis would be

\begin{enumerate}
\item{Find an equation that may be a solution to a given problem.}
\item{Evaluate the physical dimensions of each term on the left side of the equation.}
\item{Multiply the dimensions of all terms on the left hand side together to get the dimensions of the entire left hand side.}
\item{Repeat (2) and (3) for the right hand side.}
\item{Compare the dimensions of each side of the equation. If they are the same, the equation may be correct. If they are not, the equation is incorrect.}
\end{enumerate}

This game allows students to catch some mistakes in their answers. Students in our sample played dimensional analysis readily on questions that specifically asked about dimensions, for example asking which of a set of four formulas could be the surface area of an object, but also occasionally used it productively in questions aimed understanding functional relationships. For example, when asked,
\begin{quote}
Sixteen students are sharing N large cheese pizzas. Assuming that the students share the pizza evenly, which expression gives the number of students each pizza must feed?
\end{quote}
many students had difficulty choosing between the expressions $N/16$ and $16/N$, among other distractors. Two interviewees noted that the number 16 had units of students, and because the answer they were looking for had units of students, the choice must be $16/N$

Our data set was not set up to investigate the more elaborate dimensional analysis game in which students are asked to use the dimensions of relevant variables to explicitly construct formulas, or pieces of formulas, in cases where the full analytical derivation is too long, complicated, or intractable to be useful \cite{Robinett2015}, although we believe this game would be interesting to research in the future. Constructing a formula from elemental pieces, as well as understanding an incomplete formula which contains scaling information but cannot be numerically evaluated, may lead to rich student cognition.

\subsection{Estimation}

%
%
%
%
%
%
%
%
By estimation, we mean integrating personal knowledge, a corpus of memorized numbers, and approximation heuristics to obtain order-of-magnitude estimates of interesting quantities, either in physics or in everyday scenarios. Like dimensional analysis and examining extreme cases, estimation is a highly valued in the physics community and in physics education, which have a culture of ``Fermi estimates'', ``back of the envelope'' calculations, and ``order of magnitude'' estimates. For example, \textit{The Physics Teacher} publishes a ``Fermi Question'' in each issue, and several universities have undergraduate courses in estimation \cite{caltechOOM,berkeleyOOM}.

We chose to investigate estimation because performing estimates generally requires students to think about their everyday experience and find methods of quantifying it, often while building equations that multiply various such terms together. Thus, it forces students to use intuition and a formal understanding of mathematics simultaneously.

A case study by Modir et. al. \cite{Modir2014} established an estimation epistemic game involving six moves,
\begin{enumerate}
\item Problematize
\item Propose method
\item What to remember
\item See if parts are enough
\item Pure Calculations
\item Evaluation
\end{enumerate}
and documented how a student estimated the energy in a hurricane by going rapidly forward and backward between these moves.

In one of our interviews, a group of four students, Amelia, Zane, Jean, and Chris, attempt to estimate the time it would take a submersible submarine to sink to the bottom of the ocean. The group agreed to assume the ocean was 1000m deep, and Jean calculated a descent time of about fourteen seconds by assuming the sphere fell with ordinary gravitational acceleration. Several group members challenged the notion that the submersible would accelerate during its descent and proposed it would instead fall at terminal velocity, but never reached consensus before the following exchange

\begin{quote}
\begin{enumerate} 
\item Amelia: Well if you think about it based on the previous situation that we said, we said it was at a thousand meters (Jean: mmhmm) the force was two thousand newtons. Fourteen seconds technically could be legible just because a thousand meters isn't really a lot. We have a really heavy (Zane: that's true) like submersible, so it kind of makes sense in that situation.
\item Zane: let's go with it
\item Jean: go with the...
\item Zane: fourteen seconds, yeah
\item Amelia: It all depends on like, all these variables. With these variables it would make sense that it would be dropping that fast.
\item Jean: And we're assuming there's no um, buoyant force, no viscous force
\end{enumerate}
\end{quote}

Although Zane called on counterintuitions several minutes before this exchange (``it's not going to hit, you know, a hundred thousand miles per hour at the bottom.''), and repeatedly argued against the constant acceleration approach, the group decided that their calculation ``kind of makes sense'', ultimately accepting a highly unreasonable answer. Despite their incorrect conclusion, we see in this passage group members calling on a sense of whether numbers are reasonable for a given physical situation, questioning the relation between unknown parameters and quantities of interest, and examining the simplifying physical assumptions that go into their reasoning.
At the conclusion of the interview, the interviewer mentioned that their conclusion had the submersible reaching the ocean floor at roughly 300 miles per hour, and the group burst out laughing. It may be that the group's considerable efforts at sense-making failed largely due to an unfamiliarity with the relevant units, as well as neglecting to convert them into more everyday terms.

In this incident, we see a group negotiating what physical effects to model mathematically and what to ignore. This skill is essential to all physical modeling. For example, in introductory physics we often model the flight of a thrown ball using only a uniform gravitational force, giving a parabolic trajectory. In doing so, we ignore aerodynamic drag, other aerodynamic effects (e.g. lift), nonuniformity of the gravitational field, inertial forces due to Earth's rotation, magnetization of the ball in Earth's magnetic field, the Yarkovsky effect, momentum imparted by sunlight the ball absorbs, transfer of material in and out of the ball's surface, and many other effects. Some of these can be important or not for a ball, depending on the accuracy we want and the parameters of the situation. Others are effectively never important for a ball thrown on Earth, but are relevant for, e.g. dust particles in space. Physicists often estimate the sizes of such effects to see whether they belong in more complete and explicit model. By improving student estimation skills, we also empower them to build better-informed mathematical models, and to understand the extent of those models' applicability. 

\section{The Nature of Equations}\label{eqns}

In physics education, there has been considerable effort to understand the ways the different ways that students view equations epistemologically \cite{airey2009disciplinary}, e.g. whether they ought to map closely to phenomena or be treated formally, be accepted as given by authority or derived from fundamental principle, and their relationship to modeling. Here, we are interested in a different type of view of equations: their ontology, or what types of object they're considered to be.

Earlier, using the example of physicists discussing the Yukawa and Coulomb potentials, we suggested that there is a variety of ways that physicists conceive of the equations they're working with. Physicists in different contexts speaking to different audiences sometimes thought of equations as dynamic objects, with one equation transforming into another, and other times thought of them as static, with one equation being a special case of another. Additionally, when equations changed, sometimes it was the speaker or the audience actively making the change, and sometimes the equation changed without a specific agent being identified.

The three problem-solving strategies introduced so far all call on students to think about equations in new ways---to hold them accountable to common sense (estimation) and to check various features of them (dimensions and special cases). We might wonder whether interacting with equations in certain ways changes the conceptualization that students have of equations.

In watching students play epistemic games with mathematics, we saw a diversity of conceptualizations of equations emerge. For example, Alma, in working the ellipse problem, checks the special case $a = b$ with reference to the formula $A = \pi \left(\frac{a+b}{2}\right)^2$
\begin{quote}
\ldots so a plus b squared over two squared times two is four plus b that would be 2 ab. b squared plus yeah. okay. yeah. okay. so then you would have r squared plus two r squared plus r squared which equals pi four r squared over four, so I guess it's a plus b over two cause you're taking the average. Oh, it's like you're turning into a circle. that's cool. yeah.
\end{quote}
By checking the special case, the ellipse is ``turning into a circle'', but Alma makes this reference not while working with the geometric object, but with the equation and substitutions on it that she was making. In other words, the ellipse is ``turning into a circle'' in that it becomes the formula for the area of a circle when $a = b = r$. This dynamic picture of an equation mirrors that a Yukawa potential that ``becomes the Coulomb'' potential in an extreme case. She is working with the formula, but instead of saying that the formula turns into a formula for a circle, she says ``you're turning it into a circle'', referencing a geometric object (the circle) while working with a non-geometric object (the formula). We suggest that for Alma, in this moment, there is no significant distinction between the formula and the object it describes, which, if correct, shows a very strong example of binding meaning to an equation.

Similarly, Amelia was examining the equation $N(t) = N_0 e^{-t/\tau}$ for the number $N$ of particles remaining when they decay over time $t$ with a time constant $\tau$. In examining the special case where half of the original number of particles remain, Amelia described actively changing equations via procedural language, such as ``I divide each side by the initial amount. I el-en [take the natural logarithm of] each side'', but she also described changing equations not according to any fixed procedural rules, ``I changed the equation, if I'm doing this logic, because I don't remember what the half life equation is off the top of my head. So I rewrote the equation to say that $Q(t)$ is equal to one half times the initial amount times $e$ to the negative $t$. $t$ referring to just time\ldots''

In both cases, the agency in changing the equation lies in Amelia herself. In the first case, she follows formal manipulations. In the second, she is ``doing this by logic'', presumably a reference to some mix of common sense, intuition, and reasoning, as a contrast to memorization, and she created an entirely new equation based off a template from the old one, assigning specific physical meaning to each term she created.

Students can take varied stances towards the types of objects that equations are while manipulating, creating, and interpreting them in many contexts, not simply in the context of the strategies we investigated. We believe this menagerie of conceptualizations of equations and interactions with them is especially rich in these epistemic games that play out with these strategies due to their requirements to blend symbology and physical meaning.

\section{Blending and Sensemaking}

In most frameworks to analyze student use of mathematics, there is a step in which the student manipulates the equations. For example, in ACER, this step is Execution of the mathematics, described as
\begin{quote}Transforming the math structures (e.g., unevaluated integrals) in the construction component into relevant mathematical expressions (e.g., evaluated integrals) is often necessary to uncover solutions. Each mathematical tool requires a specific set of steps and basic knowledge. For example, executing a Taylor approximation may require knowledge of common expansion templates (e.g., $\sin x \approx x+ x^3/3!+\ldots$) and how to adapt these templates to the mathematical model developed previously. Alternatively, one might need to know how to compute derivatives of complex functions. The mathematical procedures performed in this component are not, at least to experts, context free. In addition to employing base mathematical skills, experts maintain awareness of the meaning of each symbol in the expression (e.g., which symbols are constants when taking derivatives).\end{quote}

Although this description indicates that the operations are not purely formal, and that the problem-solver needs to remember the context and meaning of the symbols, the steps on which we understand the equations' emergent meaning and match them to physical understanding are separate steps from the steps of symbolic manipulation under these frameworks.

Research on the manipulation step has mostly focused on the difficulties that students have in making manipulations or on the procedural resources they use while manipulating equations (for example, thinking of physically sliding a variable from the numerator of one side of an equation to the denominator of another)\cite{Wittmann2015}.

Experts use individual mathematical manipulations as sources of physical sensemaking. Kustusch et al. \cite{Kustusch2014} studied physics professors solving a thermodynamics problem that involved taking partial derivatives. There were many choices for which derivatives to take, and experts used physical insight into the derivatives' meaning to guide their choices. In a review of the literature on mathematical sensemaking inside the mathematical manipulation steps of problem-solving, Kuo et. al. found ``no studies that focused upon the mathematical processing step in quantitative problem solving or described alternatives to using equations as computational tools.''  \cite{kuo2013students} The same authors then contrasted two students, one who describes a kinematic formula in terms of its meaning via a symbolic form, another who saw the formula essentially as a black-box tool, and found that these students performed the mathematical manipulations in a problem using that kinematic concept differently. The student who understood the formula via a symbolic form was able to blend mathematical and physical reasoning to take a shortcut solution to the problem, while the other student was not. 

If we value this sort of blended sensemaking, it is valuable to find ways to encourage it in students, and we believe extreme-case reasoning is one way to do this. In order to use extreme case reasoning, students must think about formulas and physical systems simultaneously, and as a result, they find new and creative ways of conceptualizing and manipulating equations.

For example, Myra, while considering the ``springs in series'' problem, has written $\frac{T}{k_1} + \frac{T}{k_2} = \Delta L_{total} = \frac{T_{sum}}{k_{series}}$ and below it $\frac{T}{k_1} + \frac{T}{k_2} = \frac{T}{k_{series}}$ on her whiteboard, saying

\begin{quote} I'm thinking that if you apply a constant force, for k-one will give like this amount of length plus k-two will give like this amount of length, then that's like the total amount of length of the series, which equals to k over T-series. And that makes sense to me. I just don't know how you would like not put the T in the equation.
\end{quote}

Although the group did not take up her method and she soon abandoned it, Myra's expression was correct, and a short algebra step away from the desired solution. In generating this expression, Myra didn't start with basic definitions and follow a purely formal procedure. Instead, she blended her conceptual understanding of stretching with the mathematical formalism while manipulating mathematical expressions.

Shortly before, Lelia stated, "and both would contribute just like one would contribute like one would have less change than the other. they'd still both probably be a part of the stretch." Myra's key insight was to translate this ``both contributing'' intuition into a symbolic form \cite{sherin2001students}, a basic template for and equation, along with a meaning used to understand entire classes of equations that build on that template. Here, Myra uses what Sherin identifies as the ``parts of a whole'' template, $[\square + \square + \ldots]$. 

Myra fits Lelia's idea about both springs contributing stretch onto this template via the heuristic equation $\text{stretch}_1 + \text{stretch}_2 = \text{stretch}_{total}$. Then, using the definition of a spring constant, which contains a variable $\Delta L$ for the stretch of the spring, Myra substitutes in the stretch of each spring, making each term physically meaningful as she does, obtaining $\frac{T}{k_1} + \frac{T}{k_2} = \frac{T}{k_{series}}$.

In a separate instance, Bert was working on the half-Atwood problem. His solution had a sign error, $a = \frac{mg}{m-M}$ instead of the correct $a = \frac{mg}{m+M}$, due to an inconsistency in how he set up his coordinate system.

The interviewer introduced and scaffolded the extreme and special case game for Bert, who readily took it up, discovering that his solution had the blocks reversing direction based on their mass, which he rejected as intuitively incorrect. Instead of reworking the entire problem from scratch, Bert tried making small modifications to his answer to eliminate the problem, for example introducing an absolute value in the denominator to keep it from changing signs. As he continued introducing and testing new solutions, he looked at $\frac{M-m}{mg}$ as a potential solution, considered the extreme case where $M \gg m$, and said
\begin{quote}
So then this is super big that's super small. [pauses, draws a minus sign on $M$ in the numerator] Still doesn't make sense. Still not working. Cause one of these [the masses] are big then it's gonna be big acceleration. That's not what should happen. Should be as this one grows [points to $M$] it gets smaller, so like that has to be in the denominator.
\end{quote}

In suggesting that $M$ must go in the denominator, Bert has repurposed the extreme cases game. Instead of evaluating potential solutions, he is placing constraints on what the unknown correct solution must look like. Like Myra, he blends his physical intuition and symbolic forms to achieve this.

The symbol template Bert uses is a division template, $\frac{\square}{\square}$, along with a conceptual schema about inverse proportionality. It is a schema where as one quantity increases, another decreases, but in the extreme case it shows that as one quantity grows very large, another becomes very small.

In applying this symbolic form, Bert begins with his intuitive understanding that very large, heavy objects are difficult to move and blends in his formal understanding of inverse proportionality to creatively generate a new instance of the extreme case game.

Bert did not wind up solving the problem; he rejected the correct solution on the mistaken grounds that it was symmetric with respect to interchange of $m$ and $M$, but despite not coming to a complete solution, he generated unique insights as well as a partial solution by renegotiating his relationships to the equation he was searching for while playing the extreme case game.

\section{Implications for Instruction}

It is common to see backsliding in surveys of student epistemologies over the course of introductory physics. For most courses, students on average exit their college physics course with less-favorable beliefs about how to learn physics than they had when they entered. \cite{redish1998student,adams2006new} As epistemologies are tied to problem solving strategies (cite Ileana's chapter), it's likely that students' conceptions of the role of mathematics and their approaches toward using it also deteriorate over most year-long introductory sequences. This means that although we observed surprising and expert-like strategies in our problem-solving interviews, we need to be wary of the possibility that our classes lead to students using these strategies less and less with time.

The reward and feedback structures in many introductory courses focus on evaluating whether a student can perform a certain calculation correctly. This includes grades on homework and exams, and in many circumstances, the verbal feedback students receive from instructors, for example that in "initiate-response-evaluate" questioning \cite{mehan1979learning}. In most of the episodes we've cited in this chapter, students wouldn't have received positive feedback from such systems. Bert didn't get the correct answer when he found creative new applications of extreme case reasoning. Myra blended her physical intuitions with formal mathematics in a symbolic form to get an expression equivalent to the correct answer for how springs add, but her group didn't take it up, and they left the interview without have reached a consensus on the correct answer. Alma, when checking the special case of a circular ellipse, used a dynamic ontology of the equation to reinforce her understanding of the test she was performing, but wasn't able to distinguish two answers which both passed that test, and she wound up choosing the wrong answer. Each time, the students were displaying expert-like problem-solving behaviors that we might not expect to see in introductory courses, but because they didn't come to the correct final conclusion, in many classrooms they wouldn't have received points on a test, heard their teachers praise, reiterate, extend on, or dive more deeply into the reasoning, or seen their peers enthusiastically take up the same methods. Because the type of feedback students receive can significantly affect their attitude toward learning \cite{carlone2014becoming,russ2009making}, this lack of positive feedback when trying expert-like strategies could easily quench students' fledgling attempts at useful, general ways of solving problems and understanding physics.

It isn't surprising that the techniques that work for experts in problem solving are less effective for novices. Learning to use tools takes practice. Riding a bicycle is much faster and more efficient than walking once you know how to do it, but it can be wobbly, frightening, and even dangerous at first. If we want students not only to try out strategies such as testing special cases or blending intuition and formalism through symbolic forms, they need a freedom to fail, encouragement to try out new ways of thinking, and positive reinforcement when they do so. Spike and Finkelstein \cite{spike2016design}, studying recitation sections, found that the extent to which TAs do these things depends on their beliefs about the goals of instruction. When instructors expand their goals beyond seeing students perform calculations correctly (whether quantitative or qualitative) and value the growth of new and useful ways of thinking, classrooms environments can take the seeds of expert-like thought we've observed here and nurture them.

In our own courses, these observations have led us to two ways of encouraging new problem-solving behaviors. The first is asking questions which focus on evaluating the meaning of formulas, as opposed to using them as black boxes. For example, a problem from the textbook by Serway and Jewett (\cite{serway}) reads
\begin{quote}
Consider a gas at a temperature of 3500 K whose atoms can occupy only two energy levels separated by 1.5 eV \ldots Determine the ratio of the number of atoms in the higher energy level to the number in the lower energy level.
\end{quote}
The solution involves using the formula for the Boltzmann factor as a black box tool. To encourage different ways of reasoning about the formula, in a class one of us (Redish) taught recently, a quiz question asked
\begin{quote}
 When a membrane allows one kind of ion to pass through and not another, a concentration
difference can lead to an electric potential difference developing across the membrane. For example, if the concentration of NaCl on one side of a membrane is $c_1 = 10 mM$ and $c_2 =2 mM$ on the other, letting only Na+ ions through (and not Cl-) will build up a potential difference across the membrane. This is controlled by the equation that says that the electric potential energy, $q\Delta V$, balances the concentration difference effects via the Boltzmann factor thus:
$$\frac{c_1}{c_2} = e^{\frac{-q\Delta V}{k_B T}}$$
For a given set of concentrations ($c_1$ and $c_2$ fixed) would you expect increasing the
temperature to increase , decrease, or leave the Nernst potential, $\Delta V$, unaffected?
\end{quote}
This question encourages students to reason about the functional form of the Boltzmann factor, perhaps by imagining extreme cases or using symbolic forms. It also encourages students to think of $T$ not as a fixed entity, but as a parameter that can be tuned to change both the physical behavior of a system and the numerical value in an equation. 

In addition to asking questions that encourage students to reason about formulas instead of apply them in order to get the right answer, we also ask questions that encourage students to reflect on formulas without the need to extract a final correct or incorrect answer.  For example, in one of our recitation exercises, students are asked to construct their own equation to describe when a worm will begin to suffocate as we scale up its size (reducing its surface area to volume ratio) \cite{redish2013learning}. We then ask students,
\begin{quote}
Our analysis in [the previous part] was a modeling analysis. An organism like an earthworm might grow in two ways: by just getting longer or isometrically -- by scaling up all its dimensions. What can you say about the growth of an earthworm by these two methods as a result of your analysis in [the previous part]? Does a worm have a maximum size? If so, in what sense? If so, find it.
\end{quote}
These more open-ended and reflective questions ask students to use formulas - formulas they have constructed, for interpretation and coming to new inferences, both about physical systems and about the mathematical properties of equations.

Throughout this chapter, we have searched for a number of creative ways students approach problems, including thinking about the extreme cases, conceptualizing parameters in different ways, and using equations for estimation. In interviews, students do all these things, but they can easily lead the student seemingly nowhere---no correct answer to a question, no encouragement from an instructor, no adoption by peers. To encourage students to try out useful but difficult-to-master new strategies, we continue refining the way we ask questions and attend to student thinking during instruction.

\bibliographystyle{alpha}
\bibliography{sample}

\end{document}